\documentclass[a4paper]{jpconf}
\usepackage{graphicx}
\begin{document}

\newcommand{\be}{\begin{equation}}
\newcommand{\ee}{\end{equation}}
\newcommand{\Deln}{\ensuremath{\Delta N_\nu\;}}
\newcommand{\epm}{\ensuremath{e^{\pm}\;}}

\def\ie{{\it i.e.},~}
\def\eg{{\it e.g.},~}
\def\etal{{\it et al.}~}
\def\4he{$^4$He}
\def\3he{$^3$He}
\def\7li{$^7$Li}
\def\Yp{Y$_{\rm P}$~}
\def\yd{$y_{\rm D}$~}
\def\hii{H\thinspace{$\scriptstyle{\rm II}$}~}
\def\hii{H\thinspace{$\scriptstyle{\rm II}$}~}
\def\hi{H\thinspace{$\scriptstyle{\rm I}$}~}  
\def\di{D\thinspace{$\scriptstyle{\rm I}$}~}  
\def\Nnu{N$_{\nu}$~}
\def\mb{$m_{\rm B}$}
\def\nb{$n_{\rm B}$}
\def\rhob{$\rho_{\rm B}$}
\newcommand\la{\lower0.6ex\vbox{\hbox{\ensuremath{\buildrel{\textstyle<}\over{\sim}\ }}}}
\newcommand\ga{\lower0.6ex\vbox{\hbox{\ensuremath{\buildrel{\textstyle>}\over{\sim}\ }}}}

\newcommand{\obh}{\ensuremath{\Omega_{\rm B} h^2\;}}
\newcommand{\omh}{\ensuremath{\Omega_{\rm M} h^2\;}}
\newcommand{\och}{\ensuremath{\Omega_c h^2\;}}
\newcommand{\okh}{\ensuremath{\Omega_K h^2\;}}
\newcommand{\olh}{\ensuremath{\Omega_\Lambda h^2\;}}
\newcommand{\ofh}{\ensuremath{\Omega_\phi h^2\;}}
\newcommand{\omb}{\ensuremath{\Omega_{\rm B}\;}}
\newcommand{\omm}{\ensuremath{\Omega_{\rm M}\;}}
\newcommand{\omc}{\ensuremath{\Omega_c\;}}
\newcommand{\omk}{\ensuremath{\Omega_K\;}}
\newcommand{\oml}{\ensuremath{\Omega_\Lambda\;}}
\newcommand{\omf}{\ensuremath{\Omega_\phi\;}}
\newcommand{\omr}{\ensuremath{\Omega_{\rm R}\;}}

\title{Probing The Universe With Neutrinos At 20 Minutes And 400 Thousand Years}

\author{Gary Steigman}

\address{Departments of Physics and Astronomy, The Ohio State University,
191 West Woodruff Avenue, Columbus, OH 43210, USA}

\ead{steigman@mps.ohio-state.edu}

\begin{abstract}
Big Bang Nucleosynthesis (BBN) and the Cosmic Background Radiation (CBR)
provide complementary probes of the early evolution of the Universe and 
of its particle content.  Neutrinos play important roles in both cases,
influencing the primordial abundances of the nuclides produced by BBN
during the first 20 minutes as well as the spectrum of temperature
fluctuations imprinted on the CBR when the Universe is some 400 thousand
years old.  In this talk I review the physical effects of neutrinos at 
these different epochs in the evolution of the Universe and compare the 
theoretical predictions with the observational data to explore the consistency 
of the standard models of cosmology and particle physics and to constrain 
neutrino physics as well as more general, beyond-the-standard-model physics.
\end{abstract}

\section{Introduction}
The Universe is expanding and is filled with radiation.  All wavelengths 
(\eg of photons as well as the deBroglie wavelengths of freely expanding 
massive particles) are stretched along with the expansion.  As a result, 
during its earlier evolution the Universe was hot and dense.  The combination 
of high temperature (energy) and density ensures that the collision rate 
among particles is very high during very early epochs, guaranteeing that
all particles, with the possible exception of those with gravitational
strength interactions, were in equilibrium at sufficiently early times.
As the early Universe expands and cools, interaction rates decline and,
depending on the strength of their interactions, different particles fall
out of equilibrium at different epochs.  For neutrinos (the standard, 
``active" neutrinos $\nu_{e}$, $\nu_{\mu}$, $\nu_{\tau}$) the departure
from equilibrium occurs when the Universe is a few tenths of a second
old and the temperature of the CBR photons, \epm pairs, and the neutrinos 
is a few MeV.  It should be noted that departure from equilibrium is not
sharp and collisions continue to occur; for $T~\la 2-3$~MeV, the neutrino 
interaction rates become slower than the universal expansion rate 
(as measured by the Hubble parameter $H$) and the neutrinos effectively
decouple from the CBR photons and the \epm pairs present at that time. 
However, electron neutrinos (and antineutrinos) neutrinos continue to 
interact with the baryons (nucleons) via the charged-current, weak 
interactions until the Universe is a few seconds old and the temperature 
has dropped below an MeV.  Once again, this decoupling is not abrupt (the 
neutrinos do {\it not} ``freeze-out") and two body reactions among neutrons, 
protons, \epm pairs and $\nu_{e}$($\bar{\nu}_{e}$) continue to influence the 
ratio of neutrons to protons, albeit not sufficiently rapidly to permit the 
ratio to track its equilibrium value of n/p = exp$(-\Delta m/T)$, where 
$\Delta m = m_{\rm n} - m_{\rm p} = 1.29$~MeV.  As a result, the n/p ratio
decreases from $\sim 1/6$ at ``freeze-out" to $\sim 1/7$ when BBN begins at
$\sim 200$~sec ($T \approx 80$~keV).  Since the neutrinos are extremely
relativistic during these epochs, they can influence BBN in several ways.  
The standard cosmology, universal expansion rate is determined through the 
Friedman equation by the total energy density which, during these early 
epochs, is dominated by massless particles along with those massive particles 
which are extremely relativistic: CBR photons, \epm pairs, neutrinos.  The 
early Universe is ``radiation" dominated and neutrinos are a significant 
component of the ``radiation".  In addition, through their charged-current 
weak interactions the electron-type neutrinos help to control the 
neutron-to-proton ratio, effectively limiting the primordial abundance 
of \4he.  Furthermore, if there is a lepton asymmetry (\eg more $\nu_{e}$ 
than $\bar{\nu}_{e}$ or, vice-versa), much larger than the universal baryon 
asymmetry, the weak rates interconverting neutrons and protons would be 
affected as well.

Although \epm pairs annihilated during the first few seconds, the surviving
electrons (equal in number to the protons to ensure charge neutrality) and the 
CBR photons interact electromagnetically, tying the photons to the electrons 
(via Compton scattering).  Only after the electrons and nuclides (mainly 
protons and alphas) combine to form neutral atoms (``recombination") are 
the CBR photons released from the tyranny of the electrons to become freely 
propagating.  This occurs when the Universe is some 400 thousand years old 
and the relic photons, redshifted to the currently observed black body 
radiation at $T = 2.725$K, bring us a snapshot of the universe at this 
early epoch.  At this relatively late stage in the early evolution of the 
Universe, the key role of the freely propagating, relativistic neutrinos 
is in contributing to the total radiation density, determining the universal 
expansion rate (\eg the time -- temperature relation).  It should be noted 
that if the neutrino masses are sufficiently large the neutrinos will 
have become nonrelativistic and their free-streaming, at $v < c$, has 
the potential to damp density fluctuations in the baryon fluid.  This 
important topic, which is not addressed here, was covered by Scott 
Dodelson in his talk; the interested reader is directed to his 
contribution to this volume.
   
The primordial abundances of the relic nuclei produced during BBN depend on
the baryon (nucleon) density and on the early-Universe expansion rate.  The
amplitudes and angular distribution of the CBR temperature fluctuations also
depend on these same parameters (as well as on several others).  The universal 
abundance of baryons may be quantified by comparing the number of baryons 
(nucleons) to the number of CBR photons,
\be
\eta_{10} \equiv 10^{10}(n_{\rm B}/n_{\gamma}).
\ee
As the Universe expands the densities of baryons and photons decrease but 
the numbers of baryons and of CBR photons in a comoving volume are unchanged 
(post-\epm annihilation) so that $\eta_{10}$ measured at present, at BBN, and 
at recombination should all be the same.  This is one of the key cosmological 
tests.  Since the baryon mass density ($\rho_{\rm B} \equiv \Omega_{\rm B} 
\rho_{c}$, where $\rho_{c} = 3H_{0}^{2}/8\pi G$ is the present critical mass 
density) plays a direct role in the growth of perturbations, it is convenient 
to quantify the baryon abundance using a combination of $\Omega_{\rm B}$ and 
$h$, the present value of the Hubble parameter ($H_{0}$) measured in units 
of 100 kms$^{-1}$Mpc$^{-1}$, 
\be
\eta_{10} = 274~\omega_{\rm B} \equiv 274~\Omega_{\rm B}h^{2}.
\ee
The Hubble parameter, $H = H(t)$, measures the expansion rate of the Universe.
Deviations from the standard model ($H\rightarrow H'$) may be parameterized 
by an expansion rate parameter $S\equiv H'/H$.  Since in the standard model 
$H$ is determined by the energy density in relativistic particles, deviations 
from the standard cosmology ($S \neq 1$) may also be quantified by the 
``equivalent number of neutrinos" \Deln $\equiv N_{\nu} - 3$.  Prior to \epm 
annihilation, these two parameters are related by
\be
S = (1 + 7\Delta N_{\nu}/43)^{1/2}.
\ee
Note that \Deln is a convenient way to quantify {\it any} deviation from 
the standard model expansion rate and is not necessarily related to 
extra (or fewer!) neutrinos.

The question to be addressed here is, ``Are the predictions and observations 
of the baryon density and expansion rate of the Universe at 20 minutes (BBN) 
and 400 thousand years (CBR) in agreement with each other and with the 
standard models of cosmology and particle physics?".  If yes, what constraints 
are there on beyond-the-standard-model models?   Given the limited space 
allocated to this article, only the current status of this quest is summarized 
here.  The reader will find more detail and further references in my recent 
review article~\cite{bbnrev}.

\section{The Universe At 20 Minutes: BBN}
Nuclear reactions among neutrons and protons occur at large rates during the 
early evolution of the Universe but until the Universe has cooled sufficiently 
($T~\la 80$~keV, $t~\ga 3$~minutes), they fail to yield significant abundances 
of complex nuclides because of their competition with photo-destruction 
reactions involving the enormously more numerous CBR photons (gamma-rays).  
Once BBN begins the available neutrons are consumed very quickly to build 
\4he and all further nucleosynthesis among electrically charged nuclides 
(H, D, T, \3he, \4he) involves reactions which become Coulomb-supressed as the 
Universe expands and cools.  As a result BBN terminates when $T~\la 30$~keV 
($t~\ga 25$~min).  In the first $\sim 20$~minutes of its evolution the cosmic
nuclear reactor produces (in astrophysically interesting abundances) deuterium, 
helium-3 (any tritium decays to \3he), helium-4 and, because of the gaps 
at mass-5 and mass-8, only a trace amount of lithium-7 (produced mainly as 
berylium-7 which, later in the evolution captures an electron and decays to 
\7li).

\begin{figure}[h]
\begin{minipage}{18pc}
\includegraphics[width=18pc]{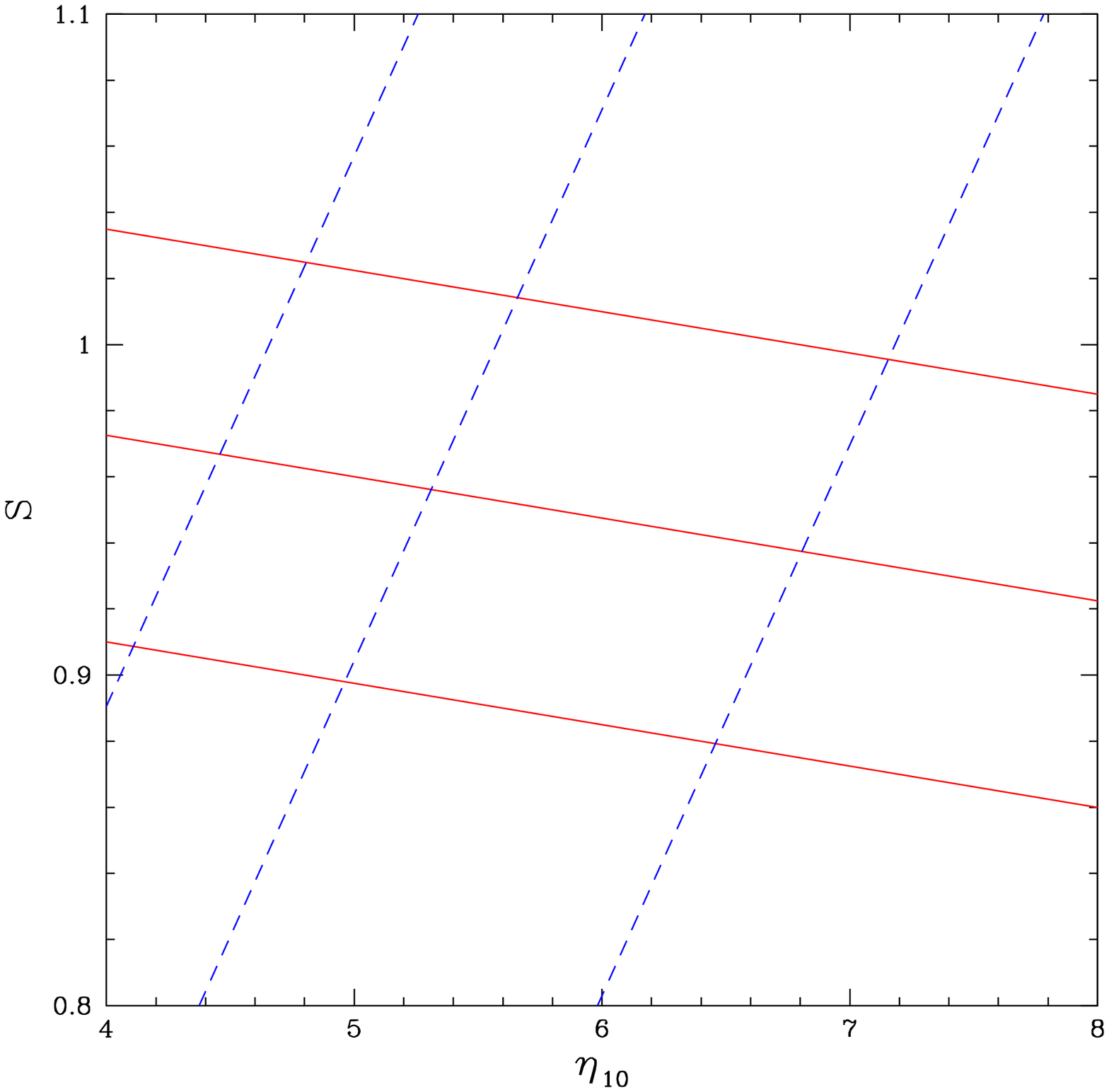}
\caption{\label{fig:svseta}Isoabundance curves for Deuterium (dashed lines) 
and Helium-4 (solid lines) in the expansion rate factor ($S$) -- baryon 
abundance ($\eta_{10}$) plane.  The \4he curves, from bottom to top, are 
for \Yp = 0.23, 0.24, 0.25.  The D curves, from left to right, are for
\yd = 4.0, 3.0, 2.0.}
\end{minipage}\hspace{2pc}%
\begin{minipage}{18pc}
\includegraphics[width=18pc]{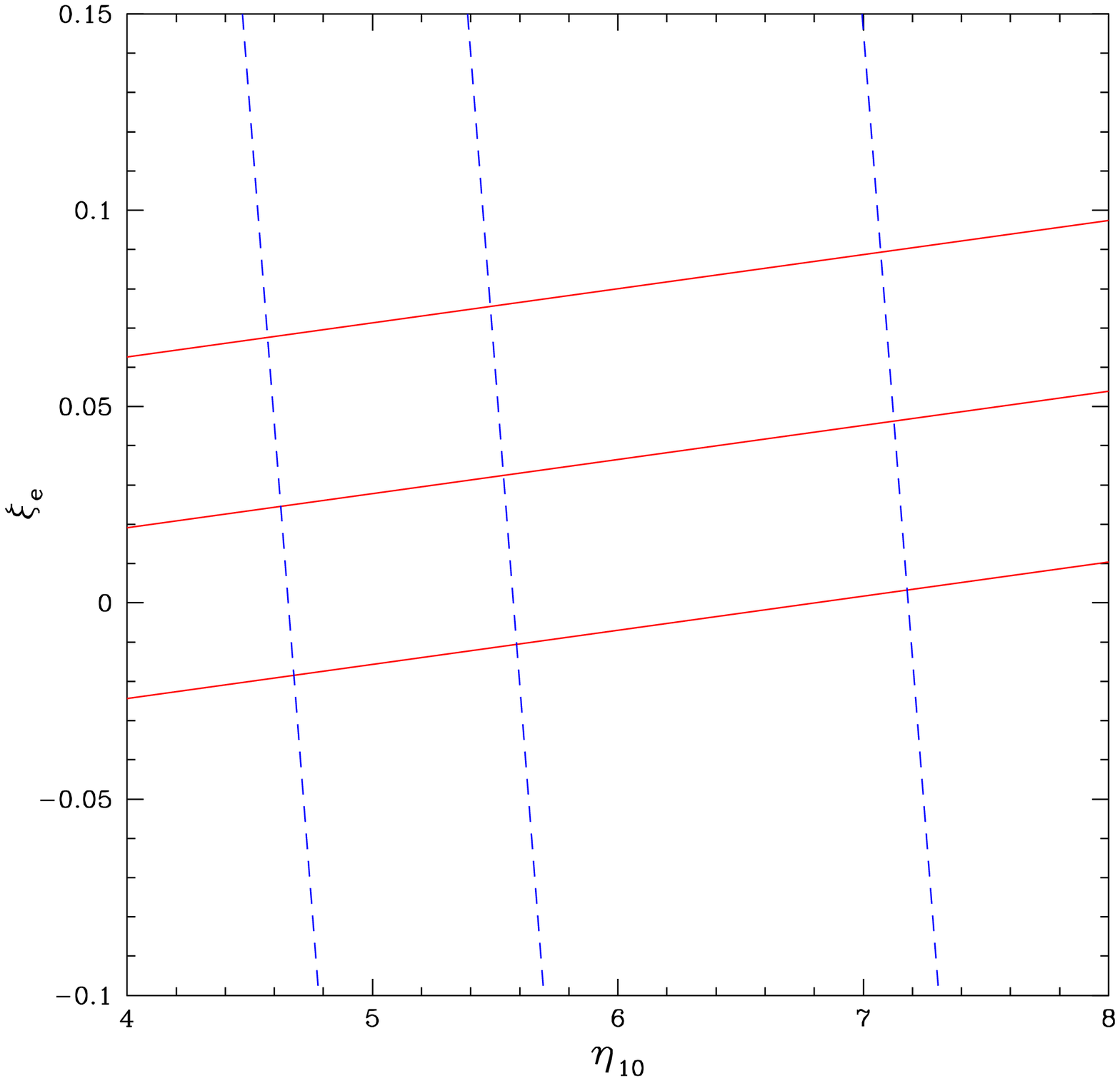}
\caption{\label{fig:xivseta}Isoabundance curves for Deuterium (dashed lines) 
and Helium-4 (solid lines) in the neutrino asymmetry ($\xi_{e}$) -- baryon 
abundance ($\eta_{10}$) plane.  The \4he curves, from bottom to top, are 
for \Yp = 0.25, 0.24, 0.23.  The D curves, from left to right, are for 
\yd = 4.0, 3.0, 2.0.}
\end{minipage} 
\end{figure}

Of the light nuclides produced during BBN, the abundances of D, \3he, and
\7li are determined by the competition between two body production and
destruction rates which are dependent on the overall density of baryons.
As a result, these nuclides are all potential baryometers.  Of these, D
is the baryometer of choice since its post-BBN evolution is simple (D is
only destroyed, burned to \3he and beyond, when gas is incorporated into
stars) and the BBN-predicted abundance is a relatively sensitive function
of the the baryon density (D/H $\propto \eta_{10}^{-1.6}$).  In contrast,
the primordial abundance of \4he is relatively insensitive to the baryon
density, depending mainly on the abundance of neutrons when BBN begins.
The \4he relic abundance is usually expressed as a ``mass fraction" \Yp 
$\equiv 4y/(1+4y)$, where $y \equiv n_{\rm He}/n_{\rm H}$ (since this 
{\it assumes} $m_{\rm He}/m_{\rm H} = 4$, \Yp is {\it not} the true helium 
mass fraction).  Since the expansion rate ($S$), in combination with the 
rate of the charged-current weak interactions, plays an important role 
in regulating the pre-BBN neutron to proton ratio, \Yp is sensitive to $S$.  
As shown by the D and \4he isoabundance curves in Figure~\ref{fig:svseta}, 
deuterium and helium-4 provide complementary probes of the universal 
baryon density and expansion rate.

While most models of particle physics beyond the standard model adopt 
(or impose) a lepton asymmetry of the same order of magnitude as the 
(very small!) baryon asymmetry ($\sim 10^{-10}\eta_{10}$), lepton 
(neutrino) asymmetries orders of magnitude larger are currently not 
excluded by any experimental data.  In analogy with $\eta_{10}$ which 
measures the baryon asymmetry, the lepton (neutrino) asymmetry, $L = 
L_{\nu} \equiv \Sigma_{\alpha} L_{\nu_{\alpha}}$, may be quantified by 
the ratios of the neutral lepton chemical potentials ($\alpha \equiv e, 
\mu, \tau$) to the temperature (in energy units) $\xi_{\nu_{\alpha}} 
\equiv \mu_{\nu_{\alpha}}/kT$, where
\be 
L_{\nu_{\alpha}} \equiv \bigg({n_{\nu_{\alpha}}-n_{\bar\nu_{\alpha}} 
\over n_{\gamma}}\bigg)= {\pi^2 \over 12 \zeta(3)}\bigg(
{T_{\nu_{\alpha}} \over T_{\gamma}}\bigg)^{3}
\bigg(\xi_{\nu_{\alpha}}+{\xi_{\nu_{\alpha}}^3 \over \pi^2}\bigg)\,. 
\label{lasym}
\ee 
Although any neutrino degeneracy always {\it increases} the energy 
density in the neutrinos, resulting in an {\it effective} \Deln$ > 0$, 
the range of $|\xi|$ of interest to BBN is limited to sufficiently small 
values that the increase in $S$ due to a non-zero $\xi$ is negligible.  
However, a small asymmetry between {\it electron} type neutrinos and 
antineutrinos ($|\xi_{e}| ~\ga 10^{-2}$; $|L| ~\ga 0.007$), while large 
compared to the baryon asymmetry, can have a significant impact on 
BBN by modifying the pre-BBN neutron to proton ratio.  The corresponding
D and \4he isoabundance curves in the $\xi_{e} - \eta_{10}$ plane are shown
in Figure~\ref{fig:xivseta}.

For restricted but interestingly large ranges of $\eta_{10}$($\omega_{\rm
B}$), $\Delta N_{\nu}$($S$), and $\xi_{e}$, Kneller and Steigman~\cite{ks} 
found simple but accurate fits to the BBN-predicted abundances of the light 
nuclides.  For D ($y_{\rm D} \equiv 10^{5}$(D/H)) and \4he (Y$_{\rm P}$), 
these are
\be
y_{\rm D} \equiv 46.5(1 \pm 0.03)\eta_{\rm D}^{-1.6}~; ~~{\rm Y}_{\rm P} \equiv 
(0.2384 \pm 0.0006) + \eta_{\rm He}/625,
\ee
where
\be
\eta_{\rm D} = \eta_{10} - 6(S-1) + {5\xi_{e} \over 4}~; ~~\eta_{\rm He} = 
\eta_{10} + 100(S-1) - {575\xi_{e} \over 4}.
\ee

\subsection{Observed Relic Abundances}
\begin{figure}[h]
\begin{minipage}{18pc}
\includegraphics[width=18pc]{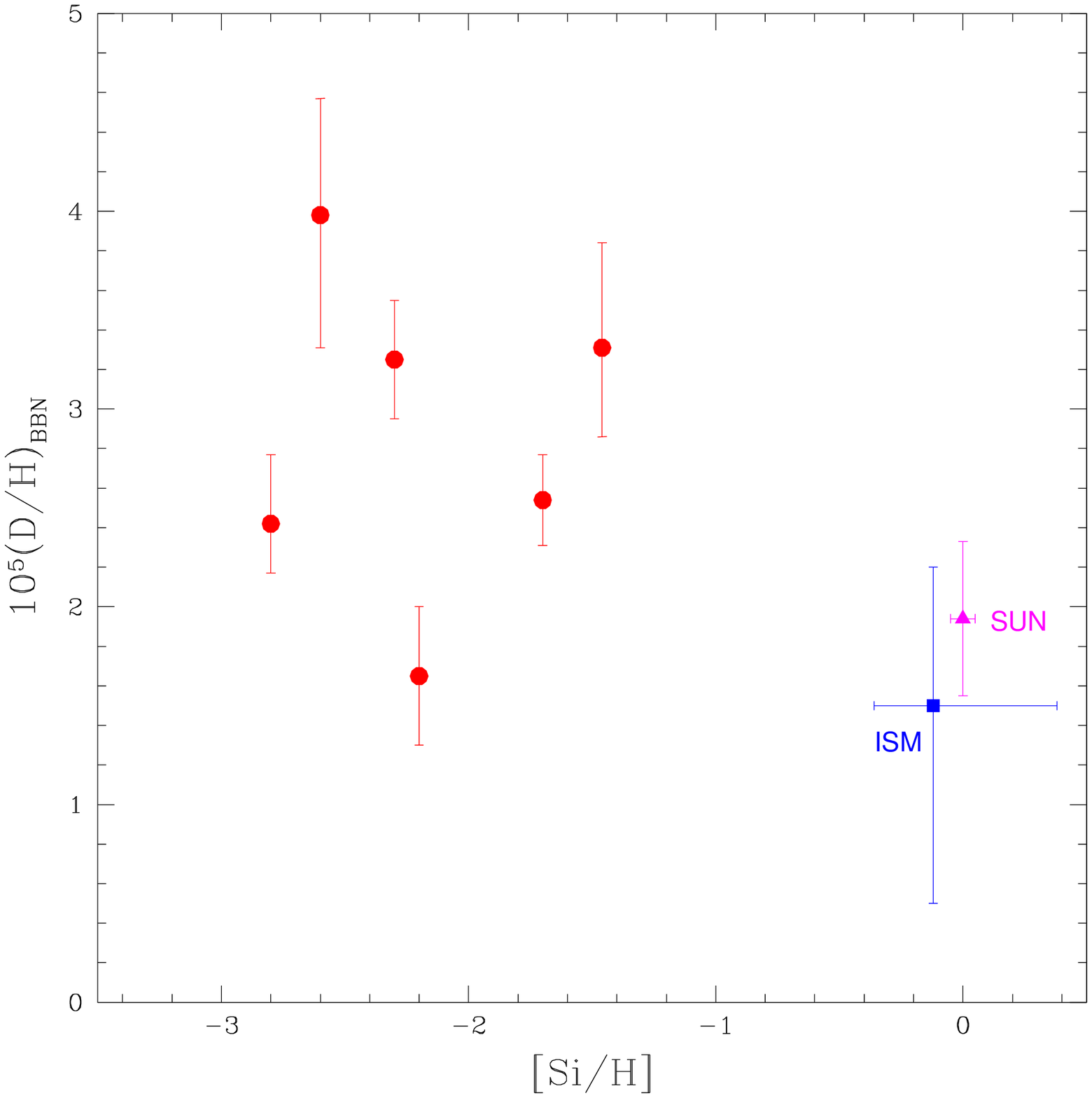}
\caption{\label{fig:dvssi06}Observationally inferred deuterium 
abundances versus metallicity for six high redshift, low 
metallicity QSOALS (filled circles).  Also shown are the 
abundances derived for the pre-solar nebula (Sun) and for 
the local interstellar medium (ISM).}
\end{minipage}\hspace{2pc}%
\begin{minipage}{18pc}
\includegraphics[width=18pc]{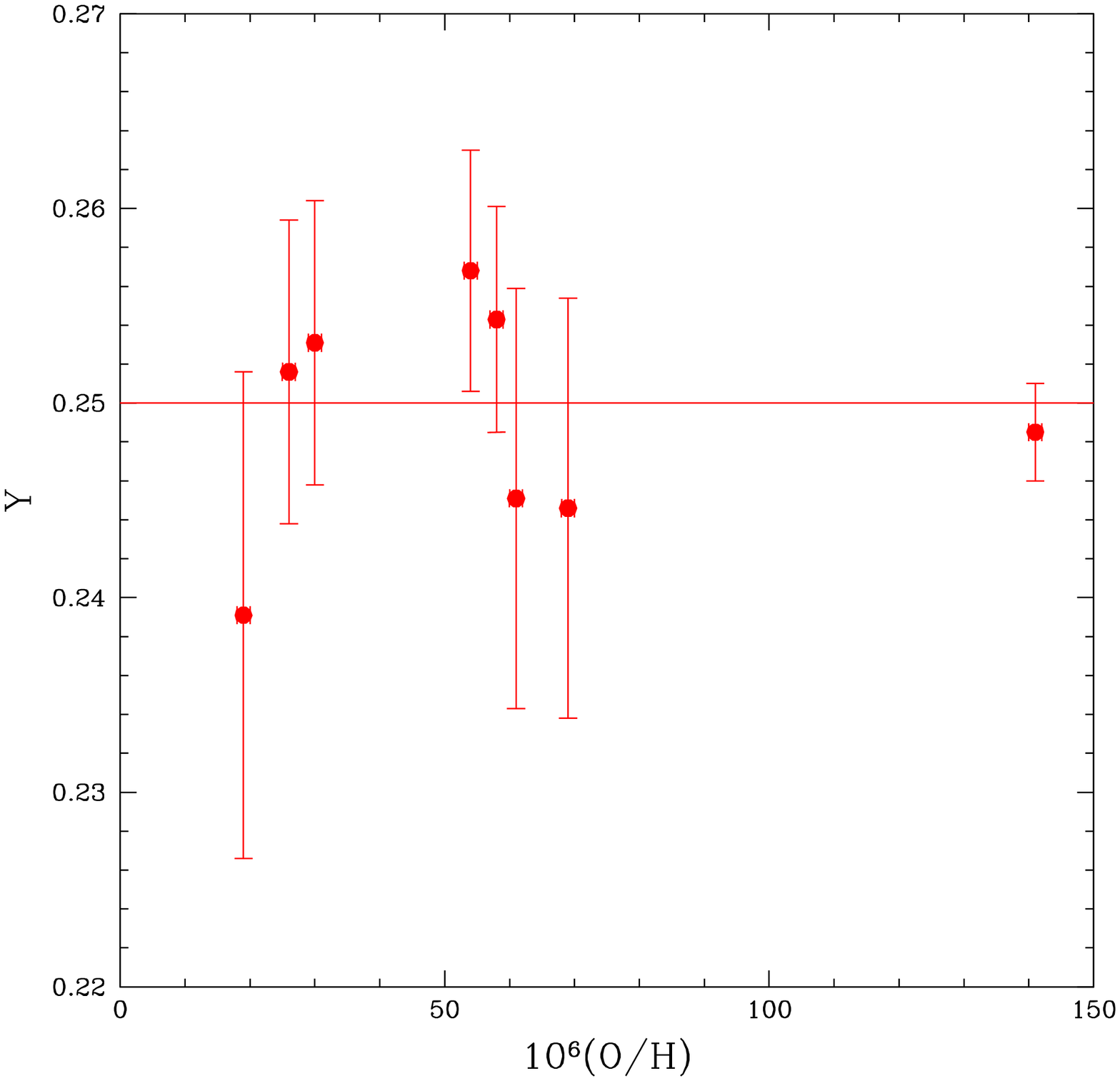}
\caption{\label{fig:hevsoos}The OS-derived \4he versus oxygen
abundances for 7 low metallicity \hii regions from IT 
and one higher metallicity \hii region from Peimbert \etal 
(filled circles).  The horizontal line shows the weighted 
mean of the 8 helium abundances.}
\end{minipage} 
\end{figure}

Although there are observations of deuterium in the solar system and the 
interstellar medium (ISM) of the Galaxy which provide interesting {\it lower} 
bounds to its primordial abundance, it is observations of relic D in a 
few (too few!), high redshift, low metallicity, QSO absorption line systems 
(QSOALS) which are of most value in enabling estimates of the primordial 
abundance.  The identical absorption spectra of \di and \hi (modulo the 
velocity/wavelength shift resulting from the heavier reduced mass of the 
deuterium atom) is a liability, limiting the number of useful targets in 
the vast Lyman-alpha forest of QSO absorption spectra (see, \eg Kirkman 
{\it et al.}~\cite{kirk} for further discussion).  Through 2003 there were 
only five QSOALS with deuterium detections leading to reasonably reliable 
abundance determinations \cite{kirk} (and references therein); these, 
including a very recent sixth determination by O'Meara \etal \cite{omeara}, 
are shown in Figure~\ref{fig:dvssi06}.  Also shown for comparison are the 
solar system and ISM D abundances.  There is clearly excessive dispersion 
among the low metallicity D abundances which tends to mask the anticipated 
primordial deuterium plateau, suggesting that systematic errors, whose 
magnitudes are hard to estimate, may have contaminated the determinations 
of at least some of the \di and/or \hi column densities.  Despite these 
concerns, the best that can be done with the present data is to identify 
the relic deuterium abundance with the weighted mean of the high-$z$, 
low-$Z$ D/H ratios: $y_{\rm D} \equiv 2.68^{+0.27}_{-0.25}$, corresponding 
to $\eta_{\rm D} = 5.95^{+0.36}_{-0.39}$ (see eq.~5).

The post-BBN evolution of \4he is also quite simple.  As gas cycles 
through generations of stars, hydrogen is burned to helium-4 (and 
beyond), increasing the \4he abundance above its primordial value.  
Therefore the present \4he mass fraction, Y$_{0}$, has received a 
significant contribution from post-BBN, stellar nucleosynthesis, 
and Y$_{0} >$~Y$_{\rm P}$.  However, since the ``metals" such as 
oxygen are produced by short-lived, massive stars and \4he is 
synthesized (to a greater or lesser extent) by all stars, at very 
low metallicity the increase in Y should lag that in, \eg O/H, so 
that as O/H $\rightarrow 0$, Y $\rightarrow$ Y$_{\rm P}$.  Therefore, 
although \4he is observed in the Sun and in Galactic \hii regions, 
the key data for inferring its primordial abundance are provided 
by observations of helium and hydrogen emission (recombination) 
lines from low-metallicity, extragalactic \hii regions.  The present 
inventory of such regions studied for their helium content exceeds 
80 (see Izotov \& Thuan (IT)~\cite{it}).  Since for such a large data 
set even modest observational errors for the individual \hii regions 
can lead to an inferred primordial abundance whose {\it formal} 
statistical uncertainty is very small, special care must be taken 
to include hitherto ignored systematic corrections and/or errors.  
It is the general consensus that the present uncertainty in \Yp is 
dominated by the latter, rather than by the former errors.  However, 
attempts to include estimates of them have often been unsystematic 
or absent.  To account for some of these uncertainties, Olive, Steigman, 
and Walker~\cite{osw} followed the Fields and Olive~\cite{fo} analysis to 
estimate a $\sim 95\%$ confidence range of $0.228 \leq {\rm Y}_{\rm P} 
\leq 0.248$ (for Y$_{\rm P} = 0.238 \pm 0.005$, this corresponds to 
$\eta_{\rm He} = -0.25 \pm 3.15$; see eq.~5).   The most systematic 
analysis to date is that of the IT data by Olive \& Skillman 2004 (OS)~\cite{os}.  
Using criteria outlined in their earlier paper~\cite{os}, OS examined the 
IT data set and decided they could apply their analysis to only 7 of the 
82 IT \hii regions.  This tiny data set, combined with its limited range 
in oxygen abundance, severely limits the statistical significance of 
the OS conclusions.  In Figure~\ref{fig:hevsoos} are shown the OS-inferred 
helium abundances from the IT data set and from one, higher metallicity 
\hii region observed by Peimbert \etal \cite{peim}.  From these eight \hii 
regions alone there is no evidence that the helium abundance is correlated 
with the oxygen abundance; the weighted mean is $<$Y$> = 0.250\pm0.002$, 
leading to a robust $\sim 2\sigma$ upper bound on the primordial helium 
abundance of \Yp $\leq 0.254$ (corresponding to $\eta_{\rm He} \leq 9.75$).

\subsection{Comparison Between BBN-Predicted And Observed Relic Abundances}
\begin{figure}[h]
\begin{minipage}{18pc}
\includegraphics[width=18pc]{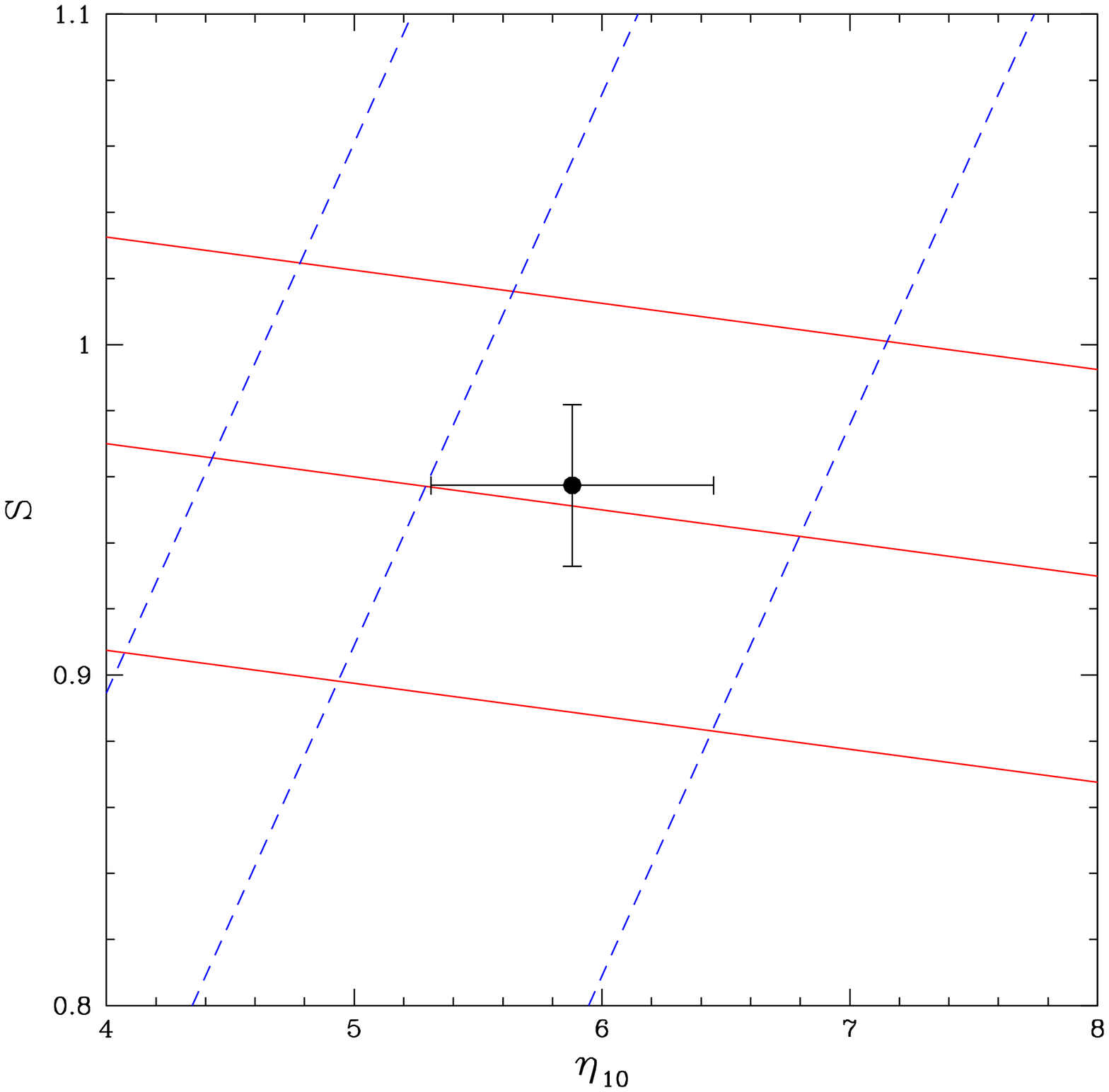}
\caption{\label{fig:svsetadhe}Isoabundance curves for Deuterium (dashed 
lines) and Helium-4 (solid lines) in the expansion rate factor -- baryon 
abundance plane (see Fig.~\ref{fig:svseta}).  The filled circle and 
error bars correspond to the adopted relic abundances.}
\end{minipage}\hspace{2pc}%
\begin{minipage}{18pc}
\includegraphics[width=18pc]{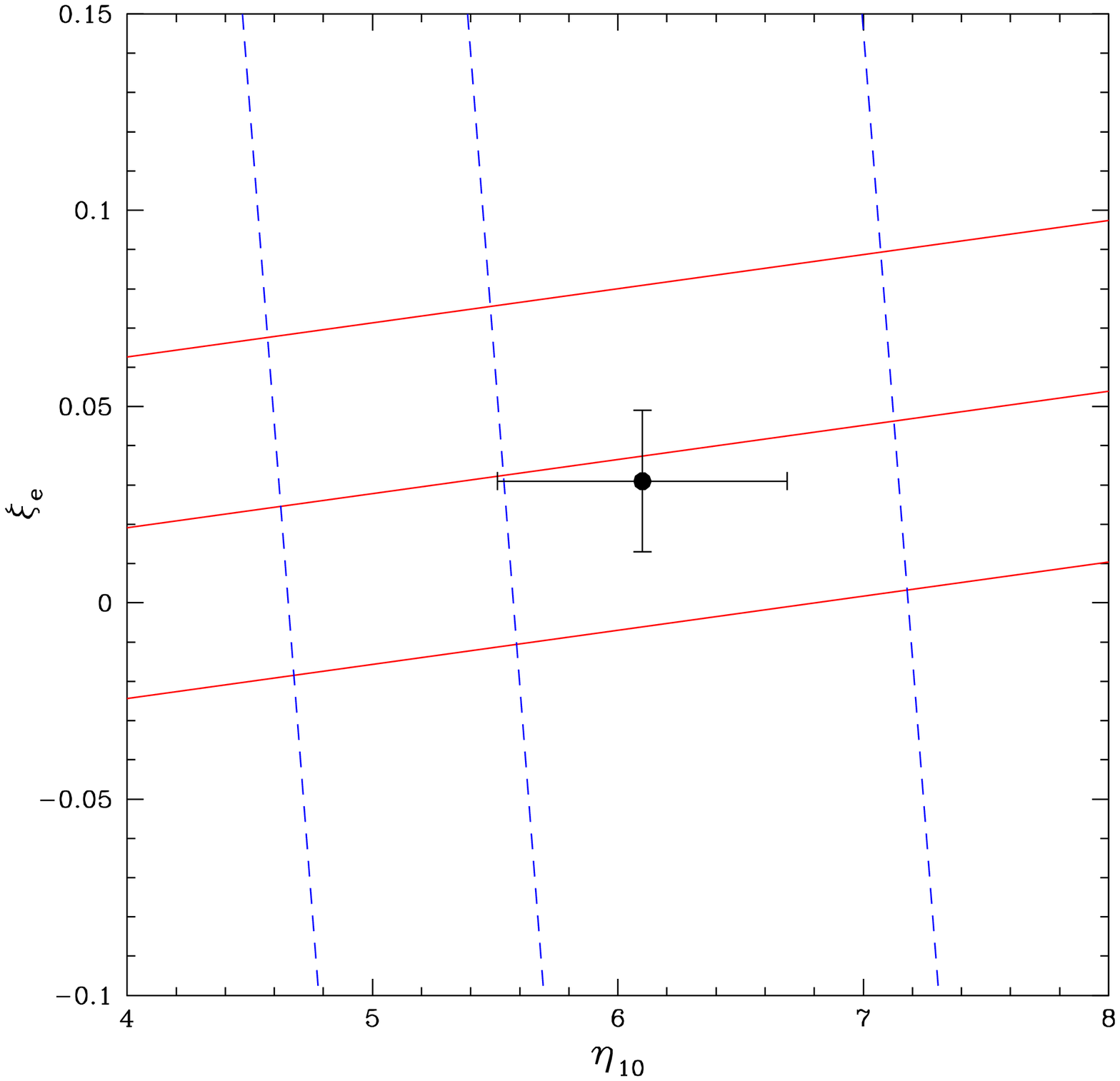}
\caption{\label{fig:xivsetadhe}Isoabundance curves for Deuterium (dashed 
lines) and Helium-4 (solid lines) in the neutrino asymmetry -- baryon 
abundance plane (see Fig.~\ref{fig:xivseta}).  The filled circle and 
error bars correspond to the adopted relic abundances.}
\end{minipage} 
\end{figure}
The relic abundances adopted here correspond to $\eta_{\rm D} = 
5.95^{+0.36}_{-0.39}$ and $\eta_{\rm He} = -0.25 \pm 3.15$.  In
the absence of a large lepton asymmetry ($\xi_{e}\ll 1$), this
implies $S = 0.942\pm0.030$ (N$_{\nu} = 2.30^{+0.35}_{-0.34}$) and 
$\eta_{10} = 5.60^{+0.38}_{-0.41}$ (\obh $= 0.0204^{+0.0014}_{-0.0015}$).  
This is consistent with the standard model ($S = 1$, \Nnu = 3) at 
$\sim 2\sigma$ (see Figure~\ref{fig:svsetadhe}).  For the standard 
expansion rate/particle content ($S = 1$, \Nnu = 3), the inferred 
values of $\eta_{\rm D}$ and $\eta_{\rm He}$ result in $\xi_{e} = 
0.043\pm 0.022$ and $\eta_{10} = 5.90^{+0.36}_{-0.39}$ (\obh $= 
0.0215^{+0.0013}_{-0.0014}$).  At $\sim 2\sigma$, this is consistent 
with no lepton asymmetry ($\xi_{e} = 0$; see Figure~\ref{fig:xivsetadhe}).  
The confrontation of the BBN predictions with the relic abundance 
observations of D and \4he reveals internal consistency (at $\la 
2\sigma$) of the standard models of particle physics (N$_{\nu} 
= 3$, $\xi_{e} = 0$) and cosmology ($S = 1$) and it fixes the baryon 
abundance to an accuracy of $\sim 7\%$ during the first few minutes 
of the evolution of the Universe.  At the same time this comparison 
sets constraints on possible deviations from these standard models
(\eg N$_{\nu} \neq 4$).  How do these BBN results compare with what 
the CBR reveals about the Universe some 400 thousand years later?

\section{Confrontation With The CBR}

The angular spectrum of CBR temperature fluctuations depends on 
several key cosmological parameters, including the baryon density 
and the relativistic energy density (for further discussion and 
references, see Hu and Dodelson 2002~\cite{hu} and Barger \etal 2003 
\cite{barger}), thereby providing a probe of $\eta_{10}$ and N$_{\nu}$ 
some 400 kyr after BBN.  However, the CBR temperature fluctuation 
spectrum is blind to any sufficiently small lepton asymmetry $|\xi_{e}| 
\ll 1$.  With \Nnu allowed to depart from the standard model value, 
Barger \etal \cite{barger} found the first year WMAP data~\cite{sperg03} 
is best fit by \obh = 0.0230 and \Nnu = 2.75, in excellent agreement 
with the purely BBN results above.  In fact, the CBR is a much 
better baryometer than it is a chronometer, so that while the 
$2\sigma$ range for the baryon density is limited to $0.0204 \leq 
\Omega_{\rm B}h^{2} \leq 0.0265$, the corresponding $2\sigma$ range 
for \Nnu was found to be $0.9 \leq {\rm N}_{\nu} \leq 8.3$~\cite{barger}. 

Quite recently the WMAP team released (and analyzed) their 3-year 
data.  For \Nnu = 3, Spergel \etal 2006~\cite{nnu} find $\Omega_{\rm B}h^{2}
= 0.0223^{+0.0007}_{-0.0009}$.  When \Nnu is free to vary, V. Simha 
and the current author, in very preliminary work in progress, find 
a similar result for the baryon density (not unexpected since in 
fitting the CBR data the baryon density and the relativistic
energy density are largely uncorrelated), $\Omega_{\rm B}h^{2} = 
0.0222\pm0.0007$, along with a $2\sigma$ range for \Nnu which is 
virtually unchanged from the previous WMAP-based result~\cite{barger}: 
$2.1 \leq {\rm N}_{\nu} \leq 8.3$.  However, this latter result, 
and those of Spergel \etal 2006 and Seljak, Slosar, and McDonald 
2006~\cite{nnu} are not in very good agreement with each other and 
until the differences are fully understood, this constraint should 
be regarded with a very large grain of salt.
 
\section{Summary}

Comparison between the BBN predictions and relic abundance observations 
of deuterium and helium-4 reveals consistency with the standard models 
of particle physics and cosmology and constrains the value of the baryon 
abundance during the first few minutes of the evolution of the Universe.  
This comparison also enables quantitative constraints on possible deviations 
from these standard models, particularly in the neutrino sector.  Some 
400 thousand years later, when the CBR photons are set free, the angular 
spectrum of temperature fluctuations encodes information about several 
key cosmological parameters, including \Nnu and the baryon density.  The 
present data reveal consistency (at $\sim 2\sigma$) between the values 
of \obh and \Nnu inferred from the first few minutes of the evolution of 
the Universe and from a snapshot of the Universe some 400 kyr later.    
While there is room for surprises, at present the standard models appear
robust.

\ack
The author's research is supported at The Ohio State University by a 
grant (DE-FG02-91ER40690) from the US Department of Energy.

\smallskip


\medskip
\begin{thebibliography}{9}
\bibitem{bbnrev} Steigman G 2006 {\it Int. J. Mod. Phys.} E {\bf 15} 1
\bibitem{ks} Kneller J P and Steigman G 2004 {\it New J. Phys.} {\bf 6} 117
\bibitem{kirk} Kirkman D, Tytler D, Suzuki N, O'Meara J and Lubin D 2003
{\it ApJS} {\bf 149}, 1
\bibitem{omeara} O'Meara J M, Burles S, Prochaska J X, Prochter G E,
Bernstein R A, and Burgess K M 2006 (astro-ph/0608302)
\bibitem{it} Izotov I T and Thuan T X 1998 {\it ApJ} {\bf 500} 188; 
Izotov I T and Thuan T X 2004 {\it ApJ} {\bf 602} 200 (IT)
\bibitem{osw} Olive K A, Steigman G, and Walker T P 2000 {\it Phys. Rep.}
{\bf 333-334} 389
\bibitem{fo} Fields B D and Olive K A 1998 {\it ApJ} {\bf 506} 177
\bibitem{os} Olive K A and Skillman E D 2001 {\it New Astron.} {\bf 6} 119; 
Olive K A and Skillman E D 2004 {\it ApJ} {\bf 617} 29 (OS)
\bibitem{peim} Peimbert M, Peimbert A and Ruiz M T 2000 {\it ApJ} {\bf 541} 
688; Peimbert A, Peimbert M and Luridiana V 2002 {\it ApJ} {\bf 565} 668
\bibitem{hu} Hu W and Dodelson S 2002 {\it Ann. Rev. Astron. \& Astrophys.}
{\bf 40} 171
\bibitem{barger} Barger V, Kneller J P, Lee H-S, Marfatia D, and Steigman
G 2003 {\it Phys. Lett. B} {\bf 569} 123
\bibitem{sperg03} Spergel D N \etal 2003 {\it ApJ Suppl.} {\bf 148} 175
\bibitem{nnu} Spergel D N \etal 2006 (astro-ph/0603449); Seljak U, Slosar 
A and McDonald P 2006 (astro-ph/0604335)

\end{thebibliography}
\end{document}